\documentclass[aps,prd,preprint,
superscriptaddress]{revtex4}

\usepackage{graphicx}
\usepackage{amsmath}
\usepackage{bm}
\usepackage{color}
\usepackage{setspace}

\allowdisplaybreaks

\def\ee{\end{eqnarray}}

\def\=:{=\hspace{-.7em}\raisebox{1.1ex}{.}\hspace{.1em}\raisebox{-0.2ex}{.} }

\def\ee{\end{eqnarray}}

\def\=:{=\hspace{-.7em}\raisebox{1.1ex}{.}\hspace{.1em}\raisebox{-0.2ex}{.} }

\newcommand {\beq}{\begin{eqnarray}}
\newcommand {\eeq}{\end{eqnarray}}

\newcommand {\1}[1]{\frac{1}{#1}}

\newcommand {\del}{\partial}

\def\dps{\displaystyle}
\def\arrayret{\vspace{5pt} \\}

\begin{document}


\title{Winding Hopfions on ${\bf R}^2 \times S^1$
}


\author{Michikazu Kobayashi}
\affiliation{
Department of Basic Science, University of Tokyo, Komaba 3-8-1, Meguro-ku, Tokyo 153-8902, Japan}
\author{Muneto Nitta}
\affiliation{Department of Physics, and Research and Education Center for Natural 
Sciences, Keio University, Hiyoshi 4-1-1, Yokohama, Kanagawa 223-8521, Japan
}



\date{\today}
\begin{abstract}
We study Hopfions 
in the Faddeev-Skyrme model 
with potential terms 
on ${\bf R}^2 \times S^1$. 
Apart from the conventional Hopfions, 
there exist winding Hopfions, that is, 
the lump (baby Skyrmion) strings with the lump charge $Q$ 
with the $U(1)$ modulus twisted $P$ times along $S^1$,  
having the Hopf charge $PQ$. 
We consider two kinds of potential terms, 
that is, the potential linear in the field 
and the ferromagnetic potential 
with two easy axes, and 
present stable solutions numerically. 
We also point out that a Q-lump carries 
the unit Hopf charge per the period in $d=2+1$.

\begin{center}
\vspace{-5cm}\hspace{5cm}
\includegraphics[width=0.5\linewidth,keepaspectratio]{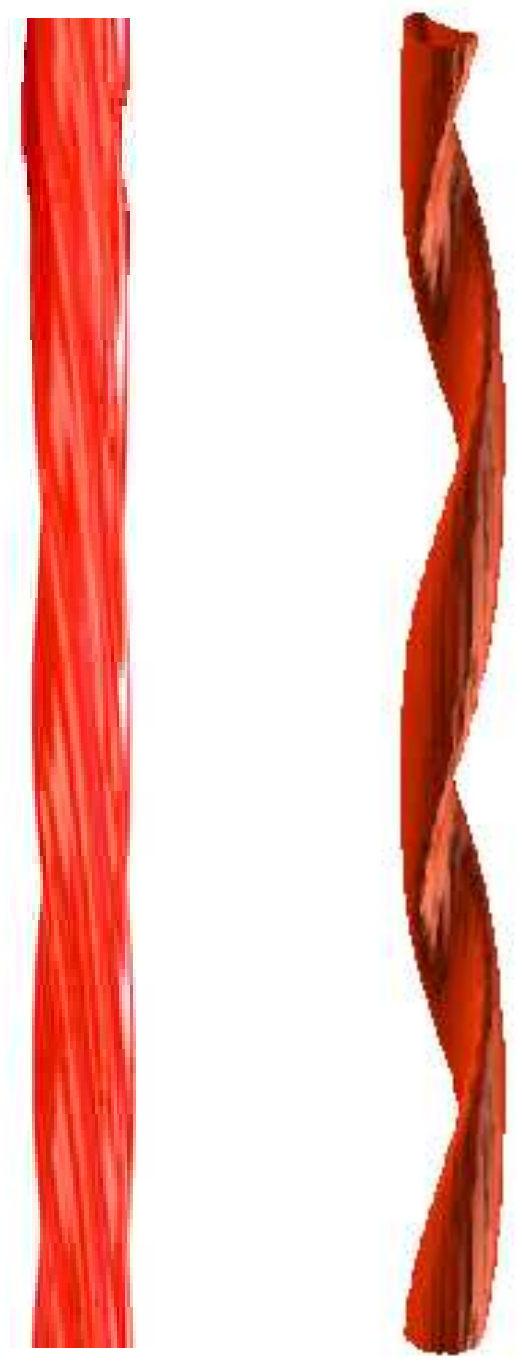}

An American liquorice 
(left) and a winding Hopfion (right).
\end{center}

\end{abstract}


\maketitle

\section{Introduction}

Knot structures are one of exotic field configurations 
with a non-trivial topology 
attracting considerable attentions in 
condensed matter physics, optics, fluid dynamics, 
and high energy physics. 
The Faddeev-Skyrme (FS) model 
\cite{Faddeev:1975,Gladikowski:1996mb,Faddeev:1996zj} 
was proposed 
as a model admitting knot structures 
as stable topological solitons  (Hopfions) 
associated with the Hopf number 
$\pi_3(S^2) \simeq {\bf Z}$  \cite{deVega:1977rk}.
It is an $O(3)$ sigma model 
with a four derivative (Skyrme) term in $d=3+1$ dimensions. 
In the early trials, one constructed 
a Hopfion as a twisted closed lump string 
\cite{deVega:1977rk,Gladikowski:1996mb}.
A lump solution associated with 
$\pi_2(S^2) \simeq {\bf Z}$ 
\cite{Polyakov:1975yp} 
is string-like in $d=3+1$.
The lump has the size and phase moduli corresponding to 
the spontaneously broken scale invariance 
the $U(1)$ rotation in 
the $n_1$-$n_2$ plane in the target space, respectively.  
When the phase modulus is twisted along the closed lump string,
it carries a Hopf number \cite{deVega:1977rk,Gladikowski:1996mb}.
Hopfions with higher charges were numerically studied, 
and, in particular, configurations with 
the Hopf charge seven or 
higher were  in fact shown to exhibit knot structures  
\cite{Battye:1998pe,Hietarinta:2000ci,Sutcliffe:2007ui}.
Hopfions in the FS model with 
potential terms 
were also studied 
\cite{Foster:2010zb,Nitta:2012kk,Harland:2013uk,Battye:2013xf,Kobayashi:2013bqa,Kobayashi:2013xoa}. 
With the ferromagnetic potential term 
\cite{Nitta:2012kk,Kobayashi:2013bqa,Kobayashi:2013xoa}, 
Hopfions are toroidal domain walls rather than knots 
if the mass is sufficiently large 
\cite{Kobayashi:2013bqa,Kobayashi:2013xoa}. 
(Un)stable Hopfions are also studied in condensed matter 
systems \cite{cond-mat} 
such as helium superfluids, 
exotic superconductors, 
ferromagnets, and 
Bose-Einstein condensates.

Other topological solitons have been studied 
in various geometries \cite{Manton:2004tk}.
In particular, flat spaces with one or more directions 
compactified as circles are useful in T-duality 
between topological solitons in different dimensions.
For instance, 
 Yang-Mills instantons 
on ${\bf R}^3 \times S^1$ and 
 $T^4$, known as Carolons,   
are related to monopoles, 
while lumps on ${\bf R} \times S^1$ \cite{Eto:2004rz} 
and $T^2$ \cite{Nakamula:2012hi} and 
vortices on ${\bf R} \times S^1$ \cite{Eto:2006mz}
and $T^2$ \cite{Eto:2007aw} 
are related to domain walls.  
Hopfions on different geometry
have not been studied much 
except for some outstanding works. 
First, Hopfions on $S^3$  were considered on 
\cite{Ward:1998pj}. 
Hopfions on general three dimensional manifolds  
were first studied in Ref.~\cite{Auckly:2005}.
Subsequently, numerical simulations for 
Hopfions on one compact direction 
${\bf R}^2 \times S^1$ and three 
compact directions $T^3$
were performed in Ref.~\cite{Jaykka:2009ry}.
For instance, when spatial infinities are identified 
for geometry ${\bf R}^2 \times S^1$ with one compact direction,  
one obtains a map from 
$S^2 \times S^1$ 
to the target space $S^2$.  
In this case, 
topology is rather complicated compared with 
conventional Hopfions on 
${\bf R}^3$ or $S^3$ classified by
$\pi_3 (S^2)\simeq {\bf Z}$; 
the map is classified by the lump charge $Q$, 
characterized by $\pi_2(S^2)\simeq {\bf Z}$, 
and the secondary invariant ${\bf Z}_{2Q}$ 
\cite{Auckly:2005,Jaykka:2009ry}. 
More physically, 
if we recall that a usual Hopfion can be made as a twisted 
closed lump string,  
we consider a lump string with the lump charge 
$Q$ winding around $S^1$ along which 
the phase modulus is twisted $P$ times. 
While such a winding Hopfion configuration carries
the Hopf charge $PQ$,
neither of the twisting $P$ nor the Hopf charge $PQ$ 
is a homotopy invariant, 
but only $Q$ and $P$ modulo $2Q$ are homotopy invariants 
\cite{Auckly:2005,Jaykka:2009ry}.  
In Refs.~\cite{Jaykka:2009ry,Hietarinta:2003vn},
winding Hopfions on ${\bf R}^2 \times S^1$ 
are shown to be dynamically unstable 
to become tangled states, 
although the Hopf charge is conserved. 
More radically, 
 the Hopf charge is not conserved anymore for 
Hopfions on $T^3$ 
which were shown to be completely unstable 
and to end up with states without zero Hopf charge.
It has been shown in Ref.~\cite{Foster:2012} that 
a straight winding Hopfion spontaneously breaks 
the rotational symmetry and 
the helical buckling instability appears 
depending on the compactification radius 
and the charge of the winding Hopfion.

A similar topological structure has been experimentally studied in superfluid $^3$He-A in a rotating container \cite{Ruutu:1994}. 
In this system, there are transverse soliton plane and 
vortices parallel
to the container intersecting the soliton plane.
Vortices wind inside the soliton plane and give
a Hopf charge.  
The topological structure has been discussed by the torus homotopy of the mapping:
$S^1 \times S^2 \to S^2$ \cite{Makhlin:1995}.

In this paper, 
we propose to  introduce potential terms
in order to stabilize winding Hopfions 
on ${\bf R}^2 \times S^1$.
The idea is that 
winding Hopfions are twisted lumps, 
and lumps are unstable to expand 
in the presence of the Skyrme term 
without potentials.
The size of the lumps is fixed in the presence of 
the Skyrme and potential terms, 
in which case the lumps are called 
baby Skyrmions in $d=2+1$ \cite{Piette:1994ug,Weidig:1998ii} 
and baby Skyrmion strings in $d=3+1$ \cite{Gisiger:1995yb}.  
The two kinds of potential terms are often considered, 
that is, a potential term 
linear in the field $V = m^2(1-n_3)$ 
\cite{Piette:1994ug,Foster:2010zb,Harland:2013uk}, 
sometimes called an old baby Skyrme potential, 
and the ferromagnetic or Ising-type potential term 
quadratic in the field 
$V = m^2 (1-n_3^2)$ 
\cite{Abraham:1992vb,Kudryavtsev:1997nw,
Nitta:2012kk,Battye:2013xf,Kobayashi:2013bqa}, 
sometimes called a new baby Skyrme potential. 
The potential term $V = m^2(1-n_3)$
admits the unique vacuum $n_3=1$, 
while the potential term $V = m^2 (1-n_3^2)$ 
admits two discrete vacua $n_3=\pm1$ 
and a domain wall solution interpolating these vacua  
\cite{Abraham:1992vb,Kudryavtsev:1997nw,
Weidig:1998ii,Nitta:2012kk,Kobayashi:2013bqa}.
With the ferromagnetic potential,  
the baby Skyrmion is in a ring shape in $d=2+1$  
\cite{Weidig:1998ii,Kobayashi:2013ju} 
and in a tube shape in $d=3+1$. 
For these potentials,
a baby Skyrmion string still has a $U(1)$ phase modulus 
while the size is fixed. 
This mode can be interpreted as a Nambu-Goldstone mode 
which is associated with 
the $U(1)$ symmetry rotating $n_1$ and $n_2$ 
spontaneously broken
in the presence of the string. 
This mode is localized  in the core of string 
and propagates along the string. 
If we consider a phase kink of the $U(1)$ modulus, 
it is unstable against the expansion 
from the Derrick's scaling argument \cite{Derrick:1964ww}.
It is then diluted along the string and eventually disappears,  
if the string extends to infinity. 
However, if the string winds around $S^1$, 
the length of the string is finite, and 
the expansion of the phase kink stops at the size of $S^1$, 
resulting in a winding Hopfion.
This configuration is similar to twisted strings \cite{Forgacs:2006pm}.

We numerically give  stable solutions of 
 $(P,Q)$ winding Hopfions, that is, 
lumps (baby Skyrmions) with the lump charge $Q$
winding around $S^1$ 
where the $U(1)$ modulus is twisted $P$ times. 
We first consider the linear potential 
$V = m^2 (1-n_3)$.  
We then consider the potential of 
the anti-ferromagnets with 
two easy axes,  given by
$V = m^2 (1-n_3^2) + \beta^2 n_1$ 
in the regime $\beta \ll m$, 
where the second term explicitly 
breaks the $U(1)$ symmetry rotating in the $n_1$-$n_2$ plane 
in the target space  
\cite{Nitta:2012xq,Nitta:2013cn,Nitta:2012wi,Kobayashi:2013ju}.
This deformation is not necessary for the stability of solutions, 
but it adds some interesting feature; 
In $d=2+1$,
there appear sine-Gordon kinks 
on a domain wall ring as a lump by this deformation  \cite{Kobayashi:2013ju}.
One unexpected feature is that 
the lump solution still posses a $U(1)$ modulus 
although the term $\beta^2 n_1$ 
explicitly 
breaks the $U(1)$ symmetry 
 in the $n_1$-$n_2$ plane. 
This $U(1)$ modulus is a Nambu-Goldstone mode 
associated with the spontaneously broken 
$U(1)$ rotation in the $x^1$-$x^2$ plane in real space. 
In fact, this solution is non-axisymmetric and 
it spontaneously breaks the rotation 
of the $x^1$-$x^2$ plane in real space.  
In $d=3+1$, sine-Gordon kinks become kink strings 
on a domain wall tube. 
In our Hopfions winding around $S^1$, 
these kink strings wrap the domain wall tube along $S^1$, 
constituting a braid. 
This configuration looks like an American liquorice.

This paper is organized as follows. 
In Sec.~\ref{sec:model}, 
after our model is explained, 
we construct baby Skyrmions 
with the two kinds of the potentials 
 in $d=2+1$ dimensions 
(they are twisted domain wall rings for $V_2$).
They can be linearly extended to baby Skyrmion strings 
(domain wall tubes for $V_2$) in $d=3+1$.
In Sec.~\ref{sec:winding-hopfion}, 
we construct winding Hopfions for the both kinds of the potentials.
In Sec.~\ref{sec:Hopf-instanton}, we point out 
that a Q-lump carries the unit Hopf charge per a unit period 
in $d=2+1$ space-time, as a Hopf instanton. 
Section \ref{sec:summary} 
is devoted to summary and discussions. 


\section{The baby Skyrme models 
and solutions\label{sec:model}}
We consider the Faddeev-Skyrme model with potential terms.
Let ${\bf n} (x)= (n_1(x), n_2(x), n_3(x))$  
be a unit three vector of scalar fields 
with a constraint ${\bf n} \cdot {\bf n} = 1$.
The Lagrangian of our model is given by
($\mu=0,1,2,3$)
\beq
&& {\cal L} = \1{2} \del_{\mu}{\bf n}\cdot \del^{\mu} {\bf n} 
 - {\cal L}_4({\bf n})
 - V({\bf n}), \quad  
 \label{eq:Lagrangian}
\eeq
with the four-derivative (Faddeev-Skyrme) term  
\beq
&& {\cal L}_4 ({\bf n})
= \kappa F_{\mu\nu}^2
= \kappa  \left[{\bf n} \cdot 
 (\partial_{\mu} {\bf n} \times \partial_{\nu} {\bf n} )\right]^2
= \kappa (\partial_{\mu} {\bf n} \times \partial_{\nu} {\bf n} )^2  
\\
&& F_{\mu\nu}={\bf n}\cdot 
(\partial_{\mu} {\bf n} \times \partial_{\nu} {\bf n} ). 
\label{eq:field-st}
\eeq 
We consider either the conventional linear potential 
\cite{Piette:1994ug,Foster:2010zb,Harland:2013uk} 
\beq
V_1({\bf n}) = m^2(1-n_3)  ,  \label{eq:pot0}
\eeq 
or the ferromagnetic potential term 
 with two easy axes  
\cite{Nitta:2012xq,Nitta:2013cn,Nitta:2012wi,Kobayashi:2013ju} 
\beq
V_2({\bf n}) = m^2(1-n_3^2) + \beta^2 n_1, 
\quad  \beta \ll m   .  \label{eq:pot}
\eeq 
The potential $V_1$ admits the unique vacuum $n_3=1$, 
and the potential $V_2$ 
admits 
the two discrete vacua
$(n_1,n_2,n_3) = (-\beta^2/2m,0, \pm \sqrt{1-(\beta^2/2m)^2})$, 
reducing to $n_3 =\pm 1$ for  $\beta = 0$.

The energy density of static configurations is  
\beq
\mathcal{E} =
\1{2} ( \del_a{\bf n}\cdot \del^a {\bf n} ) 
+ {\cal L}_4({\bf n}) + V_{1,2}({\bf n})
\eeq
with $a=1,2,3$. 

\begin{figure}
\begin{center}
\vspace{-13.5cm}
\includegraphics[width=1\linewidth,keepaspectratio]{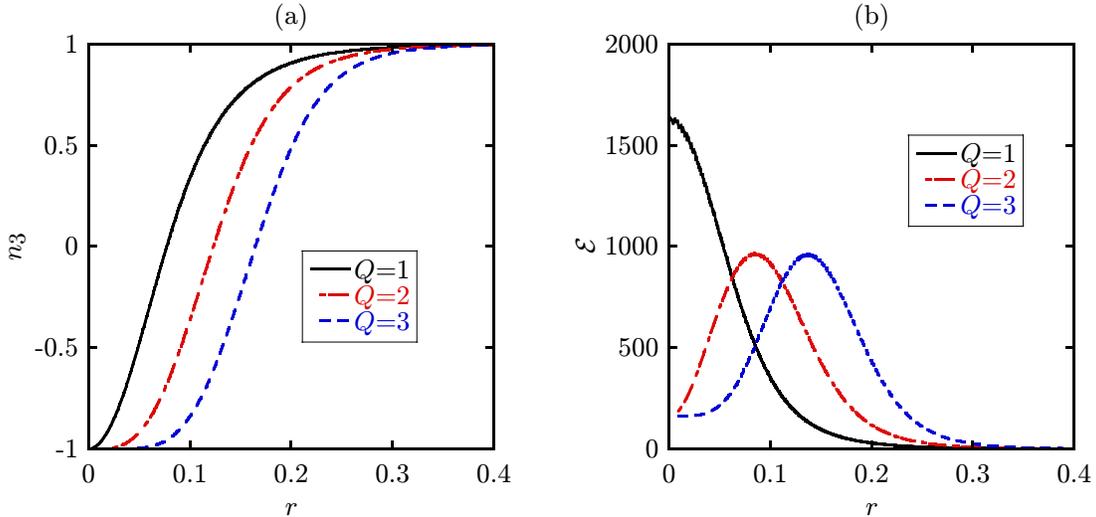}
\caption{\label{fig:lump-linear}
The $r = \sqrt{(x^1)^2 + (x^2)^2}$ dependence of (a) $n_3$ and (b) $\mathcal{E}$ of baby Skyrmions for the linear potential $V_1$ in Eq. \eqref{eq:pot0}.
We fix $\kappa = 0.002$ and $m^2 = 80$.
}
\end{center}
\end{figure}

Let us first consider the linear potential $V_1$ followed by 
the ferromagnetic potential $V_2$.
In $d=2+1$ dimensions, the static baby Skyrmion solution for the linear potential has been discussed in \cite{Piette:1994ug}. 
In order to compare the twisted string later, 
we numerically construct baby Skyrmion solutions with $Q = 1, 2, 3$, where $Q$ is the topological lump charge of $\pi_2(S^2) \simeq {\bf Z}$, given by
\beq
 Q
= \1{4 \pi}  \int d^2x\: F_{12} 
= \1{4 \pi}  \int d^2x\: {\bf n}\cdot 
(\partial_1 {\bf n} \times \partial_2 {\bf n} )
=
 \1{4 \pi}  \int d^2x\:
\epsilon_{ijk}
n_i \partial_{1} n_j  \partial_{2} n_k, \label{eq:lump-charge}
\eeq 
and plot the profiles of $n_3$ and $\mathcal{E}$ in 
Fig.~\ref{fig:lump-linear}.
Since the distributions of $\mathcal{E}$ and $n_3$ are isotropic in the $\theta = \tan^{-1}(x^2 / x^1)$-direction, we just plot the $r = \sqrt{(x^1)^2 + (x^2)^2}$-dependence of the both values.
The massive region of $n_3 \sim - 1$ localizes at the center for all $Q$. 
On the other hand, the energy density $\mathcal{E}$ 
has a hole of a small $\mathcal{E}$ region for higher $Q$.
The configuration is axisymmetric in the sense that 
the solutions is invariant under 
a combination of
the rotation of $n_1$-$n_2$ in the target space 
and the rotation in the $x_1$-$x_2$ plane in the real space, 
while the profile $n_3$ and the energy density 
$\mathcal{E}$ are invariant under either of these two transformations. 
The orthogonal combination of these $U(1)$ symmetries 
is spontaneously broken, resulting in 
the presence of a $U(1)$ Nambu-Goldstone mode.

\begin{figure}
\begin{center}
\vspace{-7cm}
\includegraphics[width=1.0\linewidth,keepaspectratio]{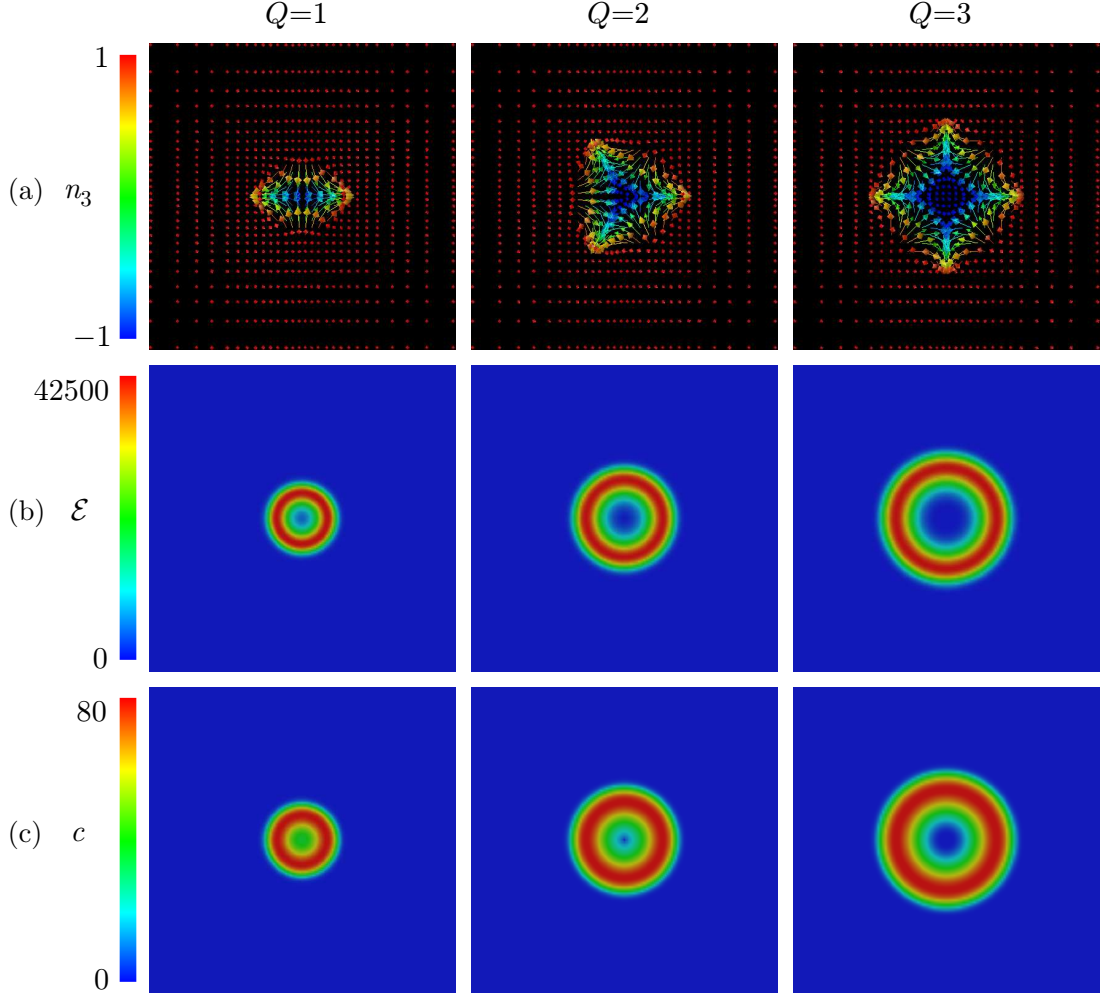}
\caption{A twisted domain wall ring as a baby Skyrmion 
 for the ferromagnetic potential $V_2$ in Eq. \eqref{eq:pot} with $\beta = 0$.
(a) The textures ${\bf n}(x)$. The color of each arrow shows the value of $n_3$.
(b) The total energy density $\mathcal{E}$. 
(c) The topological lump charge density:
$c \equiv  \{ {\bf n} \cdot (\partial_1 {\bf n} \times \partial_2 {\bf n}) \} / (4 \pi)$.
The topological charges are $Q = 1, 2, 3$ from left to right.
We fix $\kappa = 0.02$ and $m^2 = 20000$, and plot the values in the region $- 0.29 \leq x_a \leq 0.29$.
\label{fig:wall-ring0}}
\end{center}
\end{figure}
\begin{figure}[h]
\centering
\vspace{-3.0cm}
\includegraphics[width=0.9\linewidth]{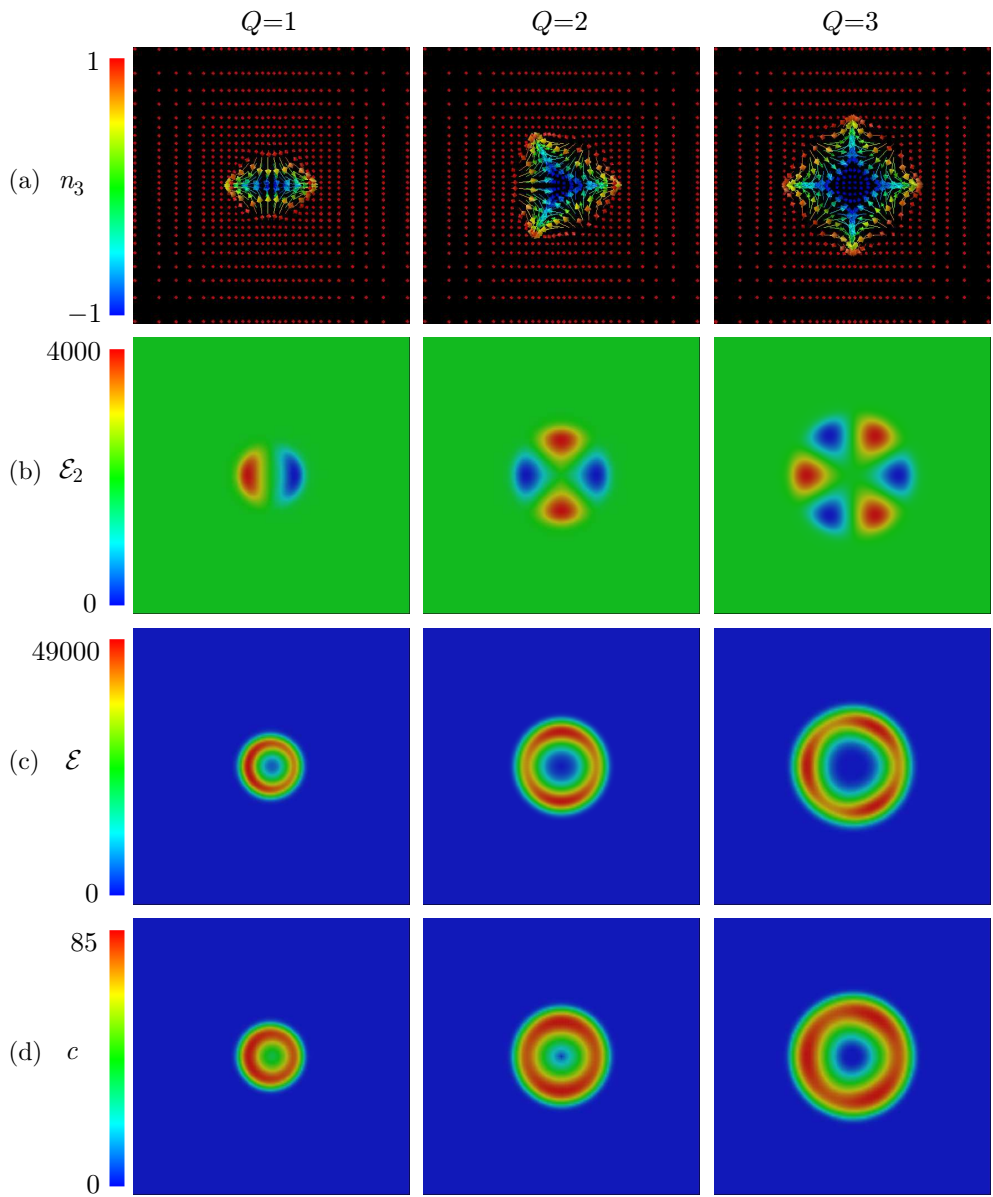}
\caption{
A twisted domain wall ring for the ferromagnetic potential $V_2$ in Eq. \eqref{eq:pot} with nonzero $\beta$.
(a) The textures ${\bf n}(x)$.
The color of each arrow shows the value of $n_3$.
(b) The energies $\mathcal{E}_2$.
(c) The total energies $\mathcal{E}$.
(d):The topological charge densities $c$.
The topological charges are $Q = 1, 2, 3$ from left to right.
We fix $\kappa = 0.02$, $m^2 = 20000$ and $\beta^2 = 2000$, and plot the values in the region $- 0.29 \leq x_a \leq 0.29$.
\label{fig:wall-ring}}
\end{figure}


Next, we consider the ferromagnetic potential $V_2$.
The potential $V_2$ with $\beta=0$ admits the 
two discrete vacua $n_3=\pm 1$ 
and a domain wall solution interpolating between them 
\cite{Abraham:1992vb,Nitta:2012wi,Kudryavtsev:1997nw}
\beq
\label{eq:wall} 
&& \theta (x^1) = 2 \arctan \exp \big\{\pm \sqrt 2 m \big(x^1 -X\big) \big\},  
  \quad 0 \leq \theta \leq \pi  , \nonumber \\
&&  n_1 = \cos \alpha \sin \theta (x^1) , \quad 
      n_2 = \sin \alpha  \sin \theta (x^1) , \quad
      n_3 = \cos \theta (x^1),
\eeq
with a phase modulus 
$\alpha$ ($0 \leq \alpha < 2\pi$) and 
the translational modulus $X \in {\bf R}$ 
of the domain wall. 
The phase modulus is a Nambu-Goldstone mode 
associated with the spontaneously broken $U(1)$ symmetry 
in the $n_1$-$n_2$ plane in the target space. 
For  $\beta \neq 0$, 
the vacua are shifted $(n_1,n_2,n_3) = (-\beta^2/2m,0, \pm \sqrt{1-(\beta^2/2m)^2})$, 
and the domain wall solutions is also modified 
for which $\alpha$ is fixed to be $\pi$. 

In $d= 2+1$ dimensions, 
a baby Skyrmion can be interpreted as 
a closed domain line for $V_2$, where 
the $U(1)$ modulus $\alpha$ 
winds $Q$ times along the wall ring 
\cite{Nitta:2012kj,Kobayashi:2013ju}.   
It carries the topological lump charge $Q$ 
defined in Eq.~\eqref{eq:lump-charge} \cite{Nitta:2012kj}.
For $\beta=0$,
 a baby Skyrmion is a twisted domain wall ring 
in Fig.~\ref{fig:wall-ring0}.
As the case of the linear potential, 
the configuration is axisymmetric 
in the sense that 
the solutions is invariant under 
a combination of
the rotation of $n_1$-$n_2$ in the target space 
and the rotation in the $x_1$-$x_2$ plane in the real space.
The Nambu-Goldstone mode appears 
as a result of the spontaneously broken $U(1)$ symmetry 
of the other combination of these two $U(1)$ symmetries. 

In the presence of nonzero $\beta$, 
the phase gradient along the domain wall ring is localized to become sine-Gordon kinks, and 
the baby Skyrmion looks like jewels on a domain wall ring 
\cite{Kobayashi:2013ju}, 
as shown in Fig.~\ref{fig:wall-ring}. 
The configurations are non-axisymmetric. 
The term $\beta^2 n_1$ explicitly breaks the 
$U(1)$ symmetry in the $n_1$-$n_2$ plane in the 
target space, unlike the cases of $V_1$ and $V_2$ 
with $\beta=0$. 
Even in the absence of the internal $U(1)$ symmetry, 
the configurations have a $U(1)$ modulus, 
which is a Nambu-Goldstone mode associated 
with the spontaneously broken 
rotational symmetry in the $x_1$-$x_2$ plane 
in real space.

For either of $V_1$ or $V_2$ or no potentials, 
one can consider 
the Hopf charge of $\pi_3(S^2)\simeq {\bf Z}$ 
in $d=3+1$,
defined by 
\beq
 C = \1{4 \pi^2}  \int d^3x\: \epsilon^{\mu\nu\rho} 
F_{\mu\nu} A_{\rho} , \label{eq:Hopf-charge}
\eeq
with a ``gauge field" $A_{\mu}$ for the field strength in 
Eq.~(\ref{eq:field-st}) satisfying $\del_{\mu}A_{\nu}-\del_{\nu}A_{\mu}=F_{\mu\nu}$ 
\cite{deVega:1977rk}. 
Hopfions carry this charge.
A preimage of a point of the $S^2$ target space 
is a closed line in real three dimensional space 
with a suitable identification at spatial infinity.
When two preimages of two arbitrary points 
on the target space make a link, 
there is a Hopf charge.
The linking number count the Hopf charge.

\section{Winding Hopfions \label{sec:winding-hopfion}}

In $d=2+1$, baby Skyrmions have been constructed 
for the linear potential $V_1$ in Fig.~\ref{fig:lump-linear}
and for the ferromagnetic potential 
with $\beta=0$ in Fig.~\ref{fig:wall-ring0}.
and with $\beta \neq 0$ in Fig.~\ref{fig:wall-ring}, 
as twisted domain wall rings. 
In $d=3+1$, the baby Skyrmion is linearly extended to 
one direction to become a baby Skyrmion string, 
which is a tube or a cylinder for $V_2$. 
We place the straight string along the $z$-direction. 
We now compactify the $z$-direction to $S^1$ 
so that the string winds around $S^1$.
 
\begin{figure}[h]
\centering
\vspace{-3.5cm}
\includegraphics[width=0.7\linewidth]{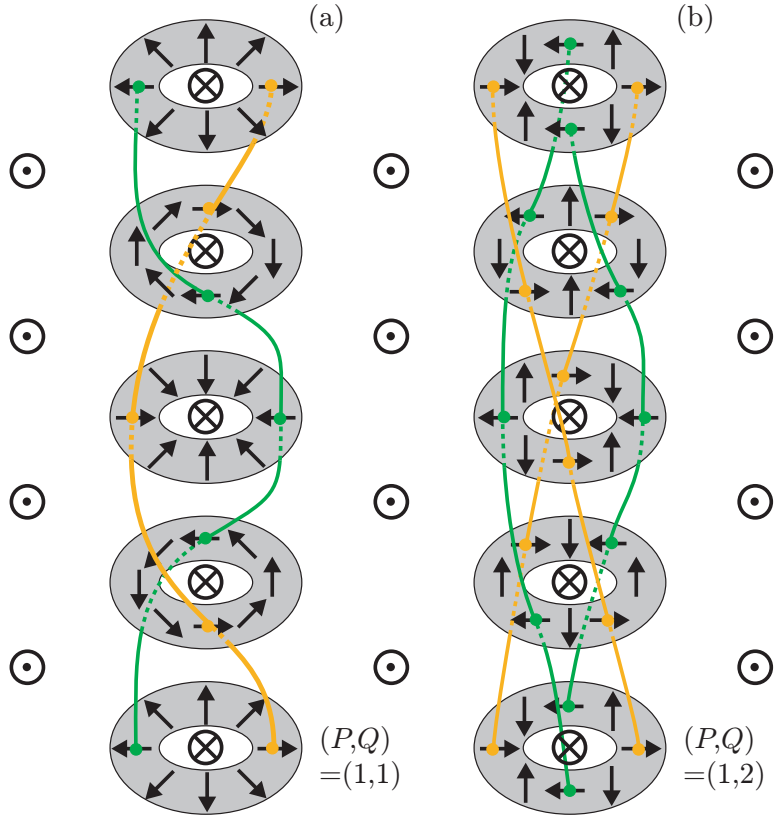}
\caption{\label{fig:cylinder} Twisted baby Skyrmion string.
The gray regions are the $n_3 = 0$ surfaces, 
which represent, for $V_2$, domain wall tubes 
that separate the two vacua  
${\bf n}=(0,0,1)$ denoted by $\odot$ and 
${\bf n}=(0,0,-1)$ denoted by $\otimes$. 
Along the $n_3 = 0$ surface, there are sequences of baby Skyrmions with the charges of (a) $Q = 1$ and  (b) $Q = 2$. From the bottom to the top, ${\bf n}$ rotates by $2 \pi P$ ($P = 1$) in the $n_1$--$n_2$ plane, defining the number of twists $P$.
The yellow and green lines indicate the locus of the $n_1 = 1$ and $n_1 = -1$ states along the tubes, respectively.
The linking numbers are $1$ and $2$ for 
the left and right figures, respectively.}
\end{figure}

The baby Skyrmion has the $U(1)$ modulus 
corresponding to the rotation around the $n_3$ axis 
in the $S^2$ target space. 
Let us rotate the $U(1)$ modulus 
clockwise along the string 
as in Fig.~\ref{fig:cylinder}.
We show the examples 
with the lump charges $Q=1$ and $2$ 
in Fig.~\ref{fig:cylinder} (a) and (b), respectively.

The spatial infinities 
in transverse directions $r = \sqrt{x^2+y^2} \to \infty$ 
are identified (one point compactification). 
Namely we consider
\beq
 {\bf R}^3 (x,y,z) \Rightarrow S^1(z) \times S^2(x,y),
\eeq
instead of $S^3$ for the usual boundary condition without 
strings.

With introducing two complex scalar fields 
$\phi^T=(\phi_1,\phi_2)$ satisfying 
$|\phi_1|^2+|\phi_2|^2=1$, 
the three-vector scalar fields $n_i$ 
can be written by the Hopf map  
\beq
  n_i = \phi^\dagger \sigma_i \phi 
\label{eq:Hopf-map}
\eeq
by using the Pauli matrices $\sigma_i$.
Note that $\phi$ parametrizes 
$S^3 \simeq SU(2)$ because of the constraint.
As an initial configuration, 
we consider an ansatz
\begin{align}
\phi = \begin{pmatrix} e^{i P g(z)} \cos\{ q_r f(r) \} \cr e^{- i q_\theta \theta} \sin\{q_r f_r(r)\} \end{pmatrix} ,\qquad
0 \leq r \leq \infty ,\quad
0 \leq \theta \leq 2\pi, \quad
0 \leq z < L_z, 
 \label{eq:ansatz}
\end{align}
with the integers $P,q_r,q_{\theta} \in {\bf Z}$ 
and the cylindrical coordinates $(r,\theta,z)$.
The monotonically decreasing function $f(r)$ satisfies
\begin{align}
f(r \to 0) \to \frac{\pi}{2}, \quad
f(r \to \infty) \to 0,
\end{align}
and the monotonically increasing function $g(r)$ satisfies
\begin{align}
g(r \to 0) \to 0, \quad
f(r \to L_z) \to 2 \pi.
\end{align}
From the Hopf map in Eq.~(\ref{eq:Hopf-map}), we have 
\begin{align}
\begin{array}{c}
\dps
n_1 = \cos\{ P g(z) - q_\theta \theta\} \sin\{2 q_r f(r)\}, \arrayret
\dps
n_2 = - \sin\{ P g(z) - q_\theta \theta\} \sin\{2 q_r f(r)\}, \quad
n_3 = \cos\{2 q_r f(r)\}.
\end{array}
\end{align}

Let us first calculate the lump charge $Q$ 
at each slice of $z=$ constant. 
We obtain 
\begin{align}
\begin{split}
\dps Q
&= \frac{1}{4 \pi} \int d^2x \: \epsilon_{ijk} n_i \partial_x n_j \partial_y n_k
= - \frac{q_r q_\theta}{2 \pi} \int_0^\infty dr\: \int_0^{2 \pi} d\theta\: \sin\{ 2 q_r f(r)\} f^\prime(r) \\
&= \frac{q_\theta \{1 - (-1)^{q_r}\}}{2}.
\end{split}
\end{align}
From this, we see that $q_r$ is not needed. 
We may take $q_r=1$ and $Q = q_\theta$, 
but we leave $q_r$ general to the end of calculation.

We now calculate the Hopf number 
$\pi_3(S^2)\simeq {\bf Z}$
for $n_i$ parameterizing $S^2$
as the Skyrme charge (baryon number) $\pi_3(S^3) \simeq {\bf Z}$ 
for $\phi$ 
parameterizing $S^3$.  
To this end, real scalar fields defined by
$\phi_1=m_1 + i m_2$, $\phi_2=m_3+im_4$ can be read
\begin{align}
\begin{array}{c}
\dps
m_1 = \cos\{q_r f(r)\} \cos\{P g(z)\}, \quad
m_2 = \cos\{q_r f(r)\} \sin\{P g(z)\}, \arrayret
\dps
m_3 = \sin\{q_r f(r)\} \cos(q_\theta \theta), \quad
m_4 = - \sin\{q_r f(r)\} \sin(q_\theta \theta).
\end{array}
\end{align}
Since the complex scalar fields $\phi$ parametrize $S^3 \simeq SU(2)$,
the Skyrmion charge (the baryon number) 
in $\pi_3(S^3)$ can be calculated from $\phi$ as 
\begin{align}
\begin{split}
C &= \frac{1}{2 \pi^2} \int d^3x\: \epsilon_{stuv} m_s \partial_1 m_t \partial_2 m_u \partial_3 m_v \\
&= - \frac{P q_r q_\theta}{4 \pi^2} \int_0^\infty dr\: \int_0^{2 \pi} d\theta\: \int_0^{L_z} dz\: \sin\{2 q_r f(r)\} f^\prime(r) g^\prime(z) \\
&= \frac{P q_\theta (1 - (-1)^{q_r})}{2} = P Q.
\end{split}
\end{align}
This precisely yields the Hopf charge in Eq.~(\ref{eq:Hopf-charge})
through the Hopf map in Eq.~(\ref{eq:Hopf-map}).

The Hopf charge can be obtained from 
the linking number of the preimages 
of two generic points, as show in Fig.~\ref{fig:cylinder}

\bigskip

\begin{figure}
\centering
\vspace{-13.5cm}
\includegraphics[width=1\linewidth]{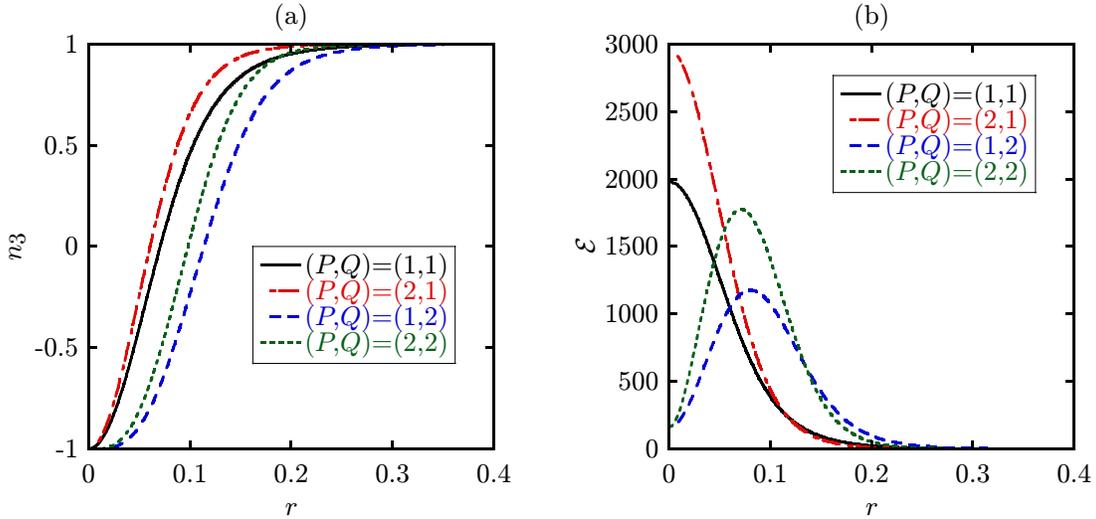}  
\caption{\label{fig:solutions-linear} 
The $r$-dependence of (a) $n_3$ and (b) $\mathcal{E}$ of winding hopfions with the linear potential $V_1$ in Eq. \eqref{eq:pot0}.
We fix $\kappa = 0.002$, $m^2 = 80$, and $L_z = 1$
}
\end{figure}
\begin{figure}
\centering
\vspace{-11.5cm}\hspace{-1cm}
\includegraphics[width=1.1\linewidth]{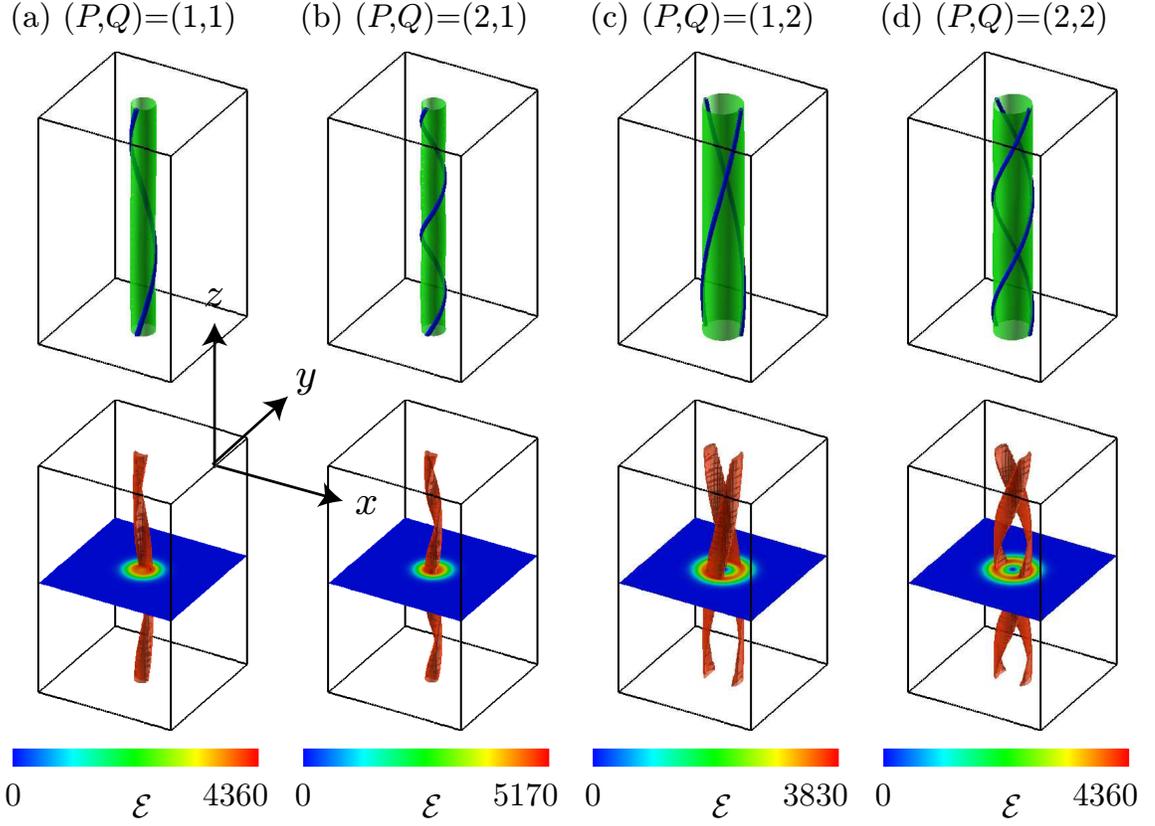}\\ 
\caption{\label{fig:solution-ising}
Winding Hopfions with the ferromagnetic potential $V_2$ in 
Eq.~\eqref{eq:pot}.
The green and blue surfaces in top panels are the isosurface of $n_3 = 0$ and $n_1 = -0.97$, respectively.
The bottom panels show $\mathcal{E}$ at the $z = L_z / 2$ plane and the isosurface of 95\% of the maximum of $\mathcal{E}$.
The topological charges are (a) $(P, Q) = (1, 1)$, (b) $(P, Q) = (2, 1)$, (c) $(P, Q) = (1, 2)$, and (d) $(P, Q) = (2, 2)$.
We fix $\kappa = 0.002$, $m^2 = 800$, $\beta^2 = 80$, and $L_z = 1$, and plot the values in the region $- 0.29 \leq x_a \leq 0.29$.
}
\end{figure}
\begin{figure}
\centering
\vspace{-12.5cm}\hspace{-1cm}
\includegraphics[width=1.1\linewidth]{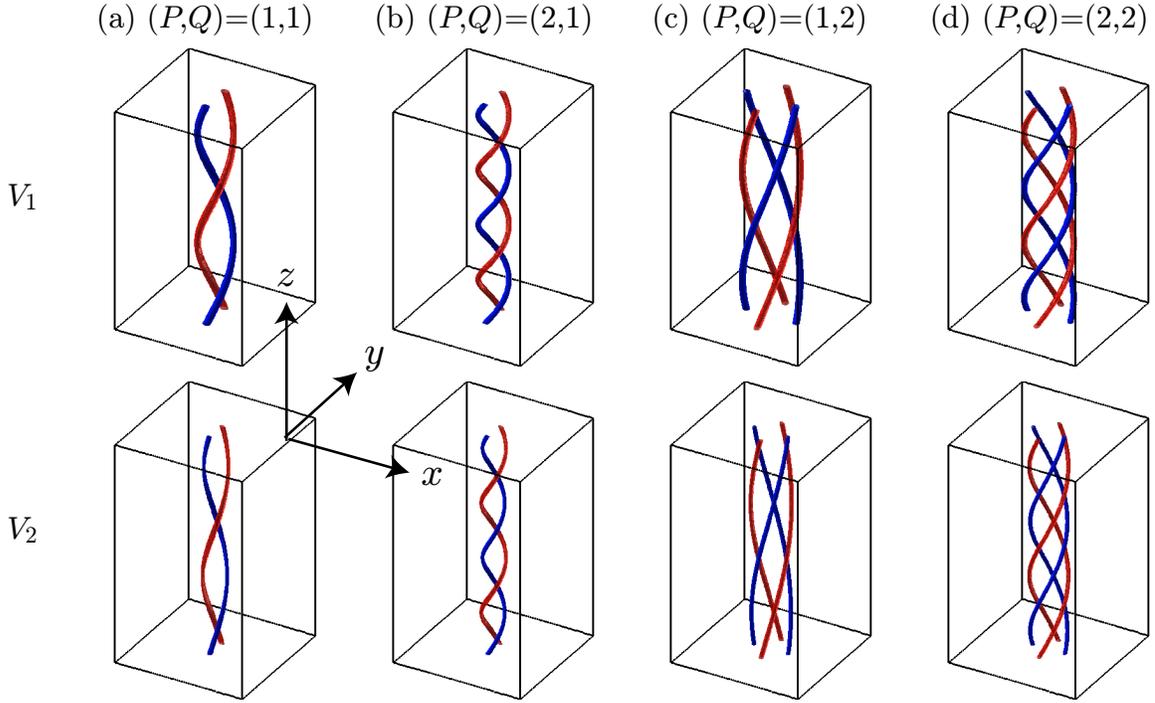}\\ 
\caption{\label{fig:solution-preimage}
The two preimages of $n_1 = 0.97$ (red surface) and $n_1 = -0.97$ (blue surface) of winding Hopfions with the linear potential $V_1$ for Fig.~\ref{fig:solutions-linear} (the upper panels) and the ferromagnetic potential $V_2$ for Fig.~\ref{fig:solution-ising} 
(the lower panels).
The topological charges are (a) $(P, Q) = (1, 1)$, (b) $(P, Q) = (2, 1)$, (c) $(P, Q) = (1, 2)$, and (d) $(P, Q) = (2, 2)$.
Linking numbers of two preimages are 1, 2, 2, and 4 for (a), (b), (c), and (d) respectively for both potentials.
}
\end{figure}

To obtain numerical solutions, 
we use the relaxation method   
starting from the ansatz 
given in Eq.~(\ref{eq:ansatz}) with $q_r=1$
as the initial configurations.
During the calculation, the topological charges do not change. 
Here, we give numerical solutions 
for $(P,Q)=(1,1), (1,2), (2,1)$ and $(2,2)$ with 
the two potentials $V_1$ in Eq.~\eqref{eq:pot0} 
and $V_2$ in Eq.~\eqref{eq:pot}.
The solutions are shown in Fig.~\ref{fig:solutions-linear} for the linear potential $V_1$.
Since the distributions of $\mathcal{E}$ and $n_3$ are isotropic in the $\theta$-direction and uniform in 
the $z$-direction, 
we just plot the $r$-dependence of the both values with fixed $z$.
Compared to the baby Skyrmion in $d = 2+1$, 
which corresponds to an unwinding Hopfion 
(untwisted string) with $P = 0$, 
the massive region of $n_3 \sim -1$ localizes more with $P$ and $\mathcal{E}$ becomes larger because $\partial_3 n_i$ becomes finite.

Our numerical solutions are shown in Fig.~\ref{fig:solution-ising} for the ferromagnetic potential $V_2$ in Eq.~\eqref{eq:pot}.
There are domain wall tubes separating 
the two vacua $n_3=\pm 1$, 
along which sine-Gordon kink strings 
localized around $n_1\sim-1$ wind.
The number of sine-Gordon kink strings is $Q$ 
at each slice of constant $z$. 
They constitute a braid structure by rotating $2\pi P /Q$ along the $z$-axis with connected to themselves or others depending on the lump charge $Q$.
Domain wall tube slightly buckles to the opposite direction of the sine-Gordon kink ($n_1 = -1$), 
because the sine-Gordon kinks pushes the domain wall tube.
The curvature of the domain wall tube becomes larger as 
larger $P$ or smaller $Q$.
This buckling of the domain wall tube is a consequence of the explicit breaking of the rotational symmetry 
in the $n^1$-$n^2$ plane due to 
the term $\beta^2 n_1$ in the potential  $V_2$.
On the other hand, we have not seen that 
the lump string buckles for the linear potential 
$V_1$ 
for our choice of the size $L_z$ of $S^1$, 
but we expect the buckling instability for a smaller $L_z$ 
even in the absence of the linear term breaking 
the rotational symmetry 
in the $n^1$-$n^2$ plane,  
as the case of Ref.~\cite{Foster:2012}.
The energy density distributions also exhibit 
braid structures as those of sine-Gordon kink strings.
The configurations are spatially twisted because of non-axisymmetry of baby Skyrmions, 
which is a unique feature of the ferromagnetic 
potential with two easy axes and 
is absent for the linear potential $V_1$
and $V_2$ with $\beta=0$.  
They look like American Liquorices.

In Fig.~\ref{fig:solution-preimage}, we plot the two preimages of $n_1 = \pm 1$ (red and blue curves, respectively) for our solutions of winding Hopfions with both linear and ferromagnetic potentials.
Two preimages make a braid structure and its linking number is consistent with the Hopf number $PQ$, being independent of two types of potential.

Now we make comments on the stability issue of these configurations. 
Unlike the case without the potential term \cite{Jaykka:2009ry,Hietarinta:2003vn}, 
our solutions are static and stable.
Since they are converged solutions 
in the relaxation method, 
they are at least at local minima of 
the configuration space.
Since we have potentials, constituent lumps are 
baby Skyrmion strings which are stable against the expansion. 
Let us move to the stability of the twisting.  
The low-energy effective theory of the $U(1)$ phase 
modulus $\alpha$ of the lump string 
can be constructed by the moduli approximation  
\cite{Manton:1981mp,Eto:2006uw}; 
one promotes the moduli to fields and integrates 
the original Lagrangian over the codimension of solitons.
We thus obtain a free $U(1)$ theory of $\alpha(t,z)$, 
a sigma model with target space $S^1$. 
Therefore the phase kink is unstable to 
expand without a potential term 
along a straight string 
from the Derrick's scaling argument \cite{Derrick:1964ww}. 
However, when the string winds around $S^1$ 
as in our case, the expansion stops 
once the kink becomes the size of $S^1$ \cite{footnote}.

\section{Hopf instantons and Q-lumps 
\label{sec:Hopf-instanton}}

Hopfions as instantons were studied in 
$d=2+1$ space-time 
\cite{Wilczek:1983cy}. 
In order to discuss an analogue of these,  
we identify the vertical direction in 
Figs.~\ref{fig:cylinder}, \ref{fig:solution-ising} 
and \ref{fig:solution-preimage} as the time direction 
in $d=2+1$ dimensions. 
Then, in our model in $d=2+1$, 
Hopf instantons can appear 
locally in time 
along a lump world-line.  
In Ref.~\cite{Wilczek:1983cy}, they studied a pair creation 
of a lump and an anti-lump 
and subsequent pair annihilation after twisting, 
corresponding to a conventional (isolated) Hopfion in Euclidean space, 
and discussed a braiding statistics. 
In our case, we have a braiding of two identical 
particles for $Q=2$ as 
in the right two figures in Fig.~\ref{fig:solution-ising}, 
and, even for $Q=1$, two points are braided 
as can be seen in Fig.~\ref{fig:solution-preimage}.

If the configuration is periodic in time 
in the model with $\beta =0$, 
it is nothing but a Q-lump.  
The Q-lump is a lump  \cite{Polyakov:1975yp}  
with the phase modulus depending linearly on time 
which is a BPS state and is stable 
\cite{Leese:1991hr}. For instance, one lump solution is 
\beq 
 u \equiv {n_1+i n_2 \over 1-n_3} = c (w - w_0) \exp (i m t)
\eeq 
with $w \equiv x^1 + ix^2$.
In the time interval with the period $m^{-1}$, 
the phase rotates once.
The configuration in this period has the unit Hopf charge 
and may be called the Hopf instanton. 
In this case, the configuration is periodic and 
the Hopf instanton is not localized. 
On the other hand, if we turn on $\beta$ in the potential, 
the Hopf instantons still exist 
as braidings mentioned above.

\section{Summary and Discussion \label{sec:summary} }
We have studied winding Hopfions 
in the Faddeev-Skyrme model on ${\bf R}^2 \times S^1$, 
with the two kinds of the potential terms, that is, 
the linear potential 
$V_1 = m^2(1-n_3)$
and 
the anti-ferromagnetic potential with two easy axes 
$V_2 = m^2 (1-n_3^2) +\beta^2 n_1$. 
The winding Hopfions are
the lump (baby Skyrmion) strings with the lump charge $Q$ 
with the $U(1)$ modulus twisted $P$ times along 
the cycle $S^1$,  
and they carry the Hopf charge $PQ$. 
We have presented stable numerical solutions 
with $P,Q= 1,2$ 
for the both type of the potentials.
The configurations are axisymmetric for the linear potential, 
and we have given the profile function of $n_3$.
The configurations are non-axisymmetric for 
the anti-ferromagnetic potential with two easy axes.  
We briefly discussed that Q-lumps in $d=2+1$ carry 
the Hopf charge density per unit period 
and can be understood as Hopf instantons. 

A winding $(P,Q)$ Hopfion carries the Hopf charge $PQ$, 
but the Hopf charge is not a homotopy invariant 
in this geometry.
The lump charge $Q$ is a homotopy invariant and 
$P$ is a homotopy invariant modulo $2Q$ 
\cite{Auckly:2005,Jaykka:2009ry}.  
Therefore, for instance, a configuration with $(P,Q)=(2,1)$ can be continuously 
deformed into a configuration with $(P,Q)=(0,1)$. 
However, our solution with $(P,Q)=(2,1)$ is at least at local minimum, implying a potential barrier to
$(P,Q)=(0,1)$.

If one makes a closed lump string with a twisted $U(1)$ phase,
one obtains an isolated Hopfion \cite{deVega:1977rk}.
Such Hopfions have been studied recently for $V_2$
with $m\neq 0, \beta=0$  \cite{Kobayashi:2013bqa}  
and $m,\beta \neq 0$ \cite{Kobayashi:2013xoa}. 
The interaction between an isolated Hopfion and 
a winding Hopfion is an important future work. 
A winding Hopfion may absorb or release 
isolated Hopfions with changing its Hopf charge.

The Faddeev-Skyrme model was proposed 
as low-energy effective theory of pure $SU(2)$ Yang-Mills theory, where Hopfions are suggested to describe glueballs 
\cite{Faddeev:1998eq}. 
Our result implies that
$SU(2)$ Yang-Mills theory on ${\bf R}^3 \times S^1$ 
may contain gluon string winding around the cycle $S^1$. 

This model admits a variety of solitons; 
in addition to a domain wall, lump (baby Skyrmion) strings 
and Hopfions \cite{Kobayashi:2013bqa} 
as elementary solitons in dimensions one, two and three, 
respectively,  
it also admits  
a composite soliton of lump strings ending on a 
domain wall (a D-brane soliton) \cite{Gauntlett:2000de,Isozumi:2004vg,Eto:2006pg}. 
The geometry ${\bf R}^2 \times S^1$ is useful to study 
duality between these solitons. 
In particular, a domain wall and a lump (baby Skyrmion) 
is T-dual to each other \cite{Eto:2004rz,Eto:2006mz}.
It will be interesting to investigate what is a dual 
object of a Hopfion.

\section*{Acknowledgements}

We thank the organizers of the conference 
``Quantized Flux in Tightly Knotted and Linked Systems," 
held in 3 - 7 December 2012 at Isaac Newton Institute 
for Mathematical Sciences, where this work was initiated. 
We would like to thank Paul Sutcliffe for useful comments. 
This work is supported in part by 
Grant-in-Aid for Scientific Research (Grant No. 22740219 (M.K.) and No. 23740198 and No. 25400268 (M.N.)) 
and the work of M. N. is also supported in part by 
the ``Topological Quantum Phenomena'' 
Grant-in-Aid for Scientific Research 
on Innovative Areas (No. 23103515 and No. 25103720)  
from the Ministry of Education, Culture, Sports, Science and Technology 
(MEXT) of Japan. 



\newcommand{\J}[4]{{\sl #1} {\bf #2} (#3) #4}
\newcommand{\andJ}[3]{{\bf #1} (#2) #3}
\newcommand{\AP}{Ann.\ Phys.\ (N.Y.)}
\newcommand{\MPL}{Mod.\ Phys.\ Lett.}
\newcommand{\NP}{Nucl.\ Phys.}
\newcommand{\PL}{Phys.\ Lett.}
\newcommand{\PR}{ Phys.\ Rev.}
\newcommand{\PRL}{Phys.\ Rev.\ Lett.}
\newcommand{\PTP}{Prog.\ Theor.\ Phys.}
\newcommand{\hep}[1]{{\tt hep-th/{#1}}}

\end{document}